# Enhanced stability of hydrogen atoms at the graphene/graphane interface of nanoribbons

Z. M. Ao,[1,*] A. D. Hernández-Nieves,[2,3,†] F. M. Peeters,[2] and S. Li[1]

1. School of Materials Science and Engineering, The University of New South Wales, Sydney, NSW 2052, Australia

2. Departement Fysica, Universiteit Antwerpen, Groenenborgerlaan 171, B-2020, Belgium

3. Centro Atomico Bariloche, 8400 San Carlos de Bariloche, Rio Negro, Argentina

**Abstract:**

The thermal stability of graphene/graphane nanoribbons (GGNRs) is investigated using density functional theory. It is found that the energy barriers for the diffusion of hydrogen atoms on the zigzag and armchair interfaces of GGNRs are 2.86 and 3.17 eV, respectively, while the diffusion barrier of an isolated H atom on pristine graphene was only ~0.3 eV. These results unambiguously demonstrate that the thermal stability of GGNRs can be enhanced significantly by increasing the hydrogen diffusion barriers through graphene/ graphane interface engineering. This may provide new insights for viable applications of GGNRs.

[*] zhimin.ao@unsw.edu.au
[†] alexande@cab.cnea.gov.ar

Graphene has been attracting enormous interests to exploit its potential applications for electronic devices due to its unique physical properties. However, such as the absence of a bandgap in the electronic spectrum of graphene and the Klein paradox as a consequence of the Dirac-type nature of the charge carriers etc., several issues have restricted the development of graphene electronics. [1,2] On the other hand, it is believed that graphene nanoribbons (GNRs) offer the possibility to achieve tuneable electronic properties. This is because their properties are highly dependent of their width and also the orientation of edges, for example, the GNRs can be turned from semiconducting to metallic by manipulating the structural parameters. [3,4] Unfortunately, to manipulate the edge structure and width of freestanding GNRs is a very challenging experimental task.[3,4] Both experimental data and the corresponding *ab initio* calculations demonstrated that the zigzag edge is metastable in vacuum due to a planar reconstruction to lower the energy of the system.[5]

Alternatively the high quality GNRs can be fabricated by selectively hydrogenating graphene or by carving GNRs on a graphane sheet. [6,7,8] But the fully hydrogenated graphene—graphane, which can be synthesized by exposing graphene to a hydrogen plasma[9] or applying a strong perpendicular electric field in the presence of a hydrogen gas,[10] is a wide-gap insulator.[11] A bandgap opening in graphene, induced by the patterned absorption of atomic hydrogen, was recently found experimentally.[5] Hybrid graphene/graphane nanoribbons (GGNRs) were also studied by *ab initio* calculations.[12,13,14] It was shown that the bandgap of GGNRs is dominated by the graphene rather than graphane,[7,12,14] and that its electronic and magnetic properties strongly depend on the degree of hydrogenation of the interface.[14] However, the hydrogen diffusion associated with high mobility of the isolated H atoms on graphene has a strong influence on the stability of the graphene/graphane interface.

In this work, we study the stability of the graphene/graphane interface in hybrid nanoribbons. We calculate the energy barrier for the diffusion of H atoms located at the graphene/graphane interface using density functional theory (DFT). All the possible diffusion pathways are analysed in order to find the minimum diffusion barrier and, therefore, to provide reference for designing the viable graphene electronic devices that possess high thermal stability in the operating conditions.

All the DFT calculations were performed using the DMOL3 code.[15] The generalized gradient approximation (GGA) with revised Perdew-Burke-Ernzerhof (RPBE) functional was employed as the exchange-correlation functional.[16] A double numerical plus polarization (DNP) was used as the basis set, while the DFT semicore pseudopotentials (DSPP) core treatment was employed for relativistic effects that replaces core electrons by a single effective potential. Spin polarization was included in all our calculations. The convergence tolerance of energy was set to $10^{-5}$ Ha (1 Ha = 27.21 eV), and the maximum allowed force and displacement were 0.02 Ha and 0.005 Å, respectively. To investigate the diffusion pathways of hydrogen atoms at the graphene/graphane interface, linear synchronous transition/quadratic synchronous transit (LST/QST)[17] and nudged elastic band (NEB)[18] tools in DMOL3 code were used, which have been well validated in order to search for the structure of the transition state (TS) and the minimum energy pathway. In the simulations, three-dimensional periodic boundary conditions were imposed, and all the atoms are allowed to relax.

The supercells used for the zigzag and armchair graphene/graphane nanoribbons are shown in Figs. 1(a) and 1(b), respectively. We minimized the interlayer interaction by allowing a vacuum width of 12 Å normal to the layer. For both type of nanoribbons, the C atoms are displaced from the C plane by about 0.29 Å due to the bonded H atoms. This value is similar to the shift of 0.32 Å that C atoms experience when a $H_2$ molecule is dissociative adsorption on graphene.[10] In both cases, this is a consequence of the change in the hybridization of the C atoms from $sp^2$ in graphene to $sp^3$ in graphane. In addition, for the zigzag GGNR both the graphene and the graphane nanoribbons are flat [see Fig. 1(a)]. However, the graphene and graphane layers are not in the same plane, they are connected with an angle of about 162° at the interface, which is consistent with previous reports.[12,14] For the armchair GGNR [Fig. 1(b)], the graphene and graphane regions are almost in the same plane, while there is little curvature in the graphene nanoribbon.

We now analyse the stability of the two types of interfaces by calculating the diffusion barriers for hydrogen atoms. For the case of a zigzag interface, there are two different types of C and H atoms, which we indicate in Fig. 1(a) as sites A and B. For the diffusion of the H atom bonded to the C atom at site A, there are two possible diffusion paths labelled as 1 and 2 in Fig. 1(a). At the site B, there are three possible diffusion pathways for the H atom that we label as 3, 4 and 5. In the case of an armchair interface, all the C atoms at the interface are equivalent from a diffusion point of view. So there are five different diffusions pathways that we label as 6-10 in Fig. 1(b). When analysing the diffusion paths, we find that all the diffusions are along linear pathways, and also that the H atom is free without directly binding to any C atom at the transition state.

The diffusion barriers for the different paths and for both types of graphene/graphane interfaces are summarized in Table I. For the zigzag interface, we found that the barriers are 3.83, 4.48, 2.86, 3.64 and 3.86 eV for the pathways 1−5, respectively. Thus, the minimum diffusion barrier for the zigzag GGNRs involves H diffusion from the carbon atom at site B along the C−C bond to its nearest carbon atom with an energy barrier of 2.86 eV. For the armchair interface, energy barriers for pathways 6−8 and 10 are 3.17, 4.07, 4.20 and 4.05 eV, respectively. The pathway number 9 involves H diffusion to the nearest C atom at site P. However, we found that this diffusion cannot occur because during the geometry optimization, the H atom at site P diffuses back to the C atom at site I. Thus, the energy barrier for H diffusion in armchair interfaces can be minimized to 3.17 eV to the second nearest C atom along Path 6.

Recently, it was reported that the diffusion barrier for a single hydrogen atom on pristine graphene layer is about 0.3 eV, which was obtained by DFT calculation using a similar method to this work.[19] Furthermore, the diffusion barriers of transition-metal (TM) atoms on graphene were reported in the range of 0.2−0.8 eV.[20] If the TM adatoms are coupled to a vacancy, the diffusion barrier would increase substantially, reaching to the range of 2.1−3.1 eV. Thus, it was claimed that adatoms with the barrier in such a magnitude are stable at room temperature,[20] supporting the notion that the stability of H atoms at GGNRs interfaces are enhanced greatly and rather stable at room temperature.

From the previous analysis, we can see that the minimum diffusion barriers for both of armchair and zigzag interfaces are about one order of magnitude larger than the energy barrier for H diffusion on pristine graphene. From Table I, we can also see that all the aforementioned H diffusion processes imply increases of several electronic Volts in the energy of the system. At the same time, this indicates that after the diffusion the energy needed for recovering the system back to the initial perfect thermodynamic state is always lower than the energy needed for distorting the interfaces. The barriers for backward diffusion are defined as the difference of energy between the final and the transition states ($E_F$-$E_T$), and can be obtained from Table I as the difference between the values of $E_F$-$E_I$ and the diffusion barrier ($E_T$-$E_I$). All the previous arguments demonstrate that the graphene/graphane interfaces are rather stable in both types of hybrid nanoribbons.

Such stability enhancement can be understood by calculating the binding energy of the H atoms in the different conditions, which is proportional to the strength of the C-H bonds. The binding energies ($E_b$) were calculated by $E_b$=$E_i$-($E_f$+$E_H$), where $E_i$ is the initial energy of the system, $E_f$ is the energy of the system after removing the H atom, and $E_H$ is the energy of an isolated H atom. For the zigzag interface, we found that the binding energy of the C-H bond at sites A and B are -4.59 and -2.80 eV, respectively. While for an H atom at site I of the armchair interface, the binding energy is -3.35 eV. All these values are larger than the binding energy of an isolated H atom on a graphene supercell containing 32 C atoms that is equal to -0.88 eV. This indicates the stability enhancement of the H atoms at graphene/graphane interfaces. The results of the binding energies also explains why for the zigzag interface it is easier to move the H atoms from site B ($E_b$=-2.80 eV) than from site A ($E_b$=-4.59 eV). This explanation is also applicable for us to understand why moving the atoms at site B ($E_b$=-2.80eV) in the zigzag interface is easier than moving the H atoms at site I ($E_b$=-3.35eV) in the armchair interface. In addition, the C-H bond length at site A (1.108 Å) is smaller than that at site B (1.112 Å). It is believed that if one bond breaks, the remaining coordinated ones would become shorter and stronger.[21,22] As shown in Fig. 1(a), the C atom at site F binds with other three C atoms, while the C atoms at both sites A and B are bonded with three C atoms and one H atom. Therefore, $E_b$ of C-C bond between sites B and F is greater than that between sites A and B. Such a strong C-C bond weakens the others bonding at site B including the C-H bond.[23] Hence, the C-H bond at site B is weaker than that at site A. The same explanation can be applicable to the case of armchair GGNR. On the other hand, C-C bond between sites R and I in Fig. 1(b) is weaker than that between sites A and B due to the effect from both C-C bonds between R and M as well as I and P. Thus, the $E_b$ of C-H bond at site I is between those at sites A and B, which is consistent with the DFT result above. Therefore, the H atom at site B can diffuse easier and the GGNR with the armchair interface is more stable than the one with the zigzag interface.

To further understand the higher stability of the H atom at site A, we analyse the atomic charges through the Mulliken method. Table II gives the atomic charges of atoms near the interfaces. We can see that atoms at both interfaces (i.e. at sites A, B, and I) are more charged than other atoms. At the interface, C atoms are more negative and the corresponding H atoms are more positive. Furthermore, it also shows that the both interfaces mainly affect the charge distribution of the first row of atoms at interfaces, while there is slight effect on the atoms of the second row at the armchair interface. This result agrees with the fact that an interface influences mainly the atoms of the first two rows.[24] It is known that the atomic charge is mostly affected by the atoms belonging to the same carbon ring, especially the nearest atoms. For the carbon and hydrogen atoms at site A, they have similar nearest atoms as sites in graphane region far apart from the interface, where the three nearest C atoms are bonded by $sp^3$ orbitals. For the C and H atoms at site B, only two nearest C atoms are bonded by $sp^3$ orbitals, the other one on its right hand side at site F is bonded by $sp^2$ orbitals. Therefore, the effect of the interface on site B is stronger than that on site A. On the other hand, for both sites A and B, there are three C atoms bonded by $sp^2$ orbitals in the carbon ring. Thus, the charge distribution of the atoms on the both sites is affected by the interface. A similar reasoning can be applied to the charge difference on the atoms at sites I and J at the armchair interface. Therefore, the C atom at site B (-0.086 $|e|$) is more chemically active than the one at site A (-0.045 $|e|$) because it has more electrons.

In summary, we studied the stability of graphene/graphane nanoribbons with both zigzag and armchair interfaces by calculating the diffusion barriers of H atoms using DFT method. We found a significantly enhanced stability of the H atoms at the graphene/graphane interfaces, if we compare it with the diffusion of an isolated hydrogen atom on pristine graphene. This is a consequence of the increase in the strength of the C-H bonds at the graphene/graphane interfaces. Our results show that both types of graphene/graphane interfaces in hybrid nanoribbons are rather stable, which increases the feasibility for future technological applications of these systems.

**Acknowledgments**

The financial supports by the Vice-Chancellor's Postdoctoral Research Fellowship Program of the University of New South Wales (SIR50/PS19184), the Flemish Science Foundation (FWO-Vl), and the Belgian Science Policy (IAP) are acknowledged. A. D. H. acknowledges also support from ANPCyT (Grant No. PICT2008-2236) and the collaborative project FWO-MINCyT (FW/08/01).

Table I. Diffusion barriers for several diffusion paths and energy differences between the states after and before the diffusion ($E_F$-$E_I$) in graphene/graphane nanoribbons.

| | Diffusion pathway | $E_F$-$E_I$ (eV) | Diffusion barrier (eV) |
|---|---|---|---|
| Zigzag interface | 1 | 1.77 | 3.83 |
| | 2 | 3.92 | 4.48 |
| | 3 | 1.90 | 2.86 |
| | 4 | 1.47 | 3.64 |
| | 5 | 1.17 | 3.86 |
| Armchair interface | 6 | 1.39 | 3.17 |
| | 7 | 2.48 | 4.07 |
| | 8 | 1.58 | 4.20 |
| | 9[a] | | |
| | 10 | 1.09 | 4.05 |

[a] We found that this diffusion path cannot occur.

Table II. Charges of C and H atoms at different sites on the graphene-graphane nanoribbons with different interfaces. The location of the sites is shown in Fig. 1, and the unit of charge is $|e|$.

| | Atom Site | C atom | H atom |
|---|---|---|---|
| Zigzag interface | A | -0.045 | 0.045 |
| | B | -0.086 | 0.057 |
| | C | -0.031 | 0.033 |
| | D | -0.030 | 0.033 |
| | E | -0.030 | 0.031 |
| | F | 0.009 | |
| | G | 0.019 | |
| | H | 0.009 | |
| Armchair interface | I | -0.087 | 0.063 |
| | J | -0.042 | 0.038 |
| | K | -0.028 | 0.033 |
| | L | -0.029 | 0.031 |
| | M | 0.021 | |
| | N | 0.005 | |
| | O | 0.004 | |
| | P | 0.021 | |
| | Q | 0.021 | |

## Figure Captions

FIG. 1. (Coulor on line) Atomic structure of graphene/graphane nanoribbons with (a) zigzag and (b) armchair interfaces after relaxation. The arrows indicate the different diffusion pathways considered. The gray and white spheres are C and H atoms, respectively.

FIG. 1

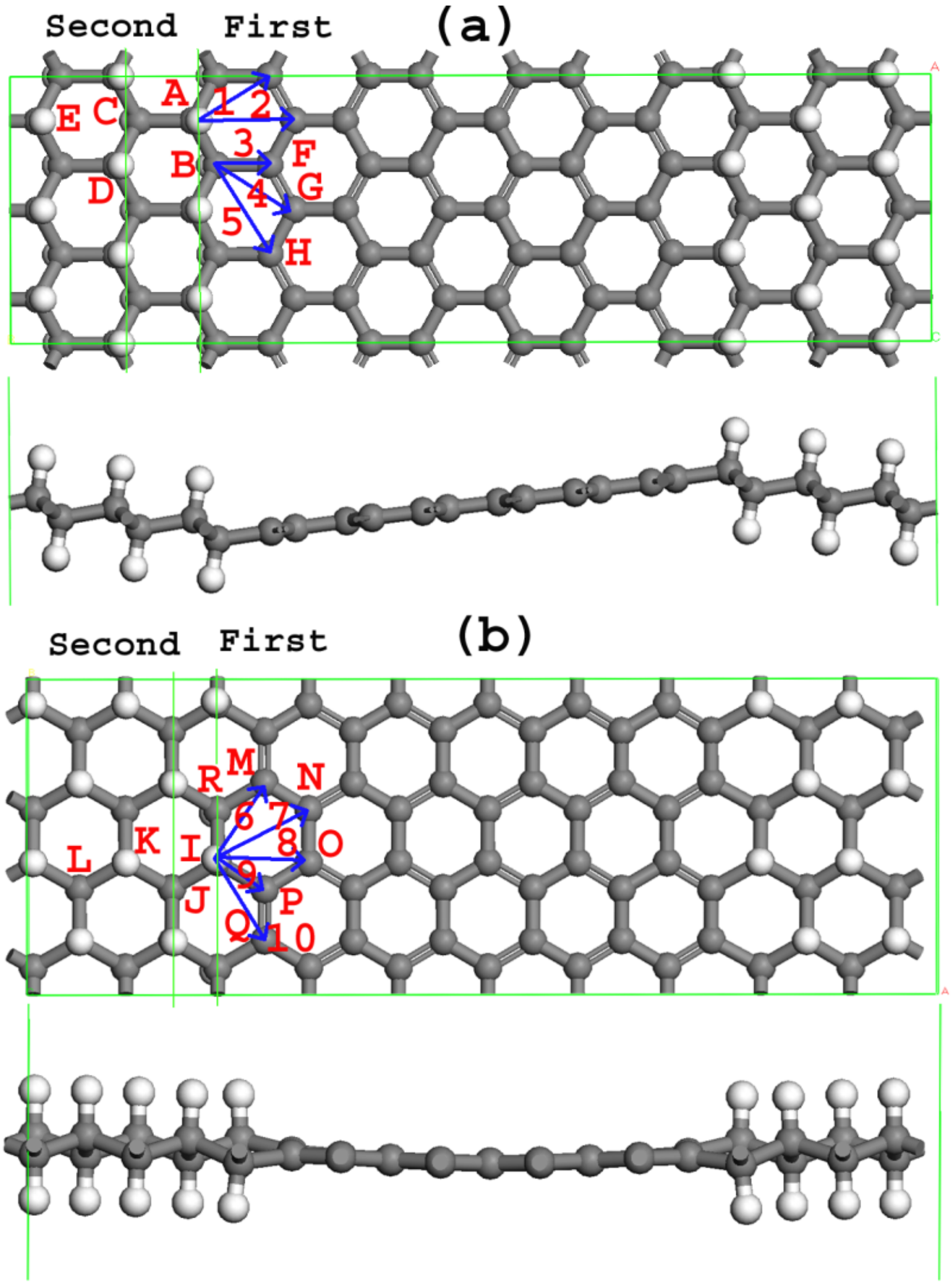

Second
First
(a)
A
1
2
E
C
B
3
F
4
G
D
5
H
Second
First
(b)
R
M
N
6
7
8
O
L
K
I
9
J
P
Q
10